\documentstyle[twocolumn,prl,aps,graphicx]{revtex}
\begin{document}
\title{Multi-mode dynamics of a coupled ultracold atomic-molecular
system}
\author{Krzysztof G\'{o}ral$^{1,3}$, Mariusz Gajda$^{2,3}$ and Kazimierz
Rz\c{a}\.{z}ewski$^{1,3}$}
\address{$^1$ Center for Theoretical Physics, $^2$ Institute of
Physics, $^3$ College of Science,\\ Polish Academy of Sciences, Aleja
Lotnik\'ow 32/46, 02-668 Warsaw, Poland}

\maketitle

\begin{abstract}
%We analyze the coherent multi-mode dynamics of a coupled
%atomic-molecular system confined in a box at ultralow
%temperatures. Starting from an atomic Bose-Einstein condensate
%with a small thermal component, we observe a complete depletion of
%the atomic and molecular condensate modes on a short time scale
%due to significant population of excited states. Contrary to the
%earlier predictions, giant coherent oscillations between the two
%condensates for typical parameters are almost completely
%suppressed. This is solely due to a weak coupling to many modes
%which leads to an effective randomization of the coherent
%evolution. Our results cast serious doubts on the common use of
%the 2-mode model for description of present experiments with
%coupled ultracold atomic-molecular systems. More importantly, they
%indicate a new destructive mechanism which will have to be
%considered when planning future experiments with ultracold
%molecules.

We analyze the coherent multi-mode dynamics of a system of coupled
atomic and molecular Bose gases. Starting from an atomic Bose-Einstein
condensate with a small thermal component, we observe a complete depletion
of the atomic and molecular condensate modes on a short time scale
due to significant population of excited states. Giant coherent
oscillations between the two condensates for typical parameters are almost
completely suppressed. Our results cast serious doubts on the common use
of the 2-mode model for description of coupled ultracold atomic-molecular
systems and should be considered when planning future experiments with
ultracold molecules.

\end{abstract}

After experimental achievement of quantum degeneracy in atomic
gases of bosons \cite{BEC_exp} and fermions \cite{fermions},
several leading groups started the quest for creation of an
analogous state in a gas of ultracold molecules \cite{molecules}.
Such systems present a potentially much more challenging
experimental task as there are several new degrees of freedom to
be controlled. A major step towards this goal has been made by the
Austin group of Heinzen \cite{Heinzen}, who photoassociated
atoms in an atomic condensate into a selected internal molecular
state producing for the first time molecules in the nK regime. Cold
molecules are believed to be created as a transient state in
experiments with Feshbach resonances as well \cite{Feshbach}. In
this novel area there are only a few theoretical works analyzing
the process of making ultracold molecules in atomic condensates
and predicting several interesting phenomena in the mixed
atomic-molecular systems. Javanainen {\it et al.} in a series of
papers \cite{Juha} analyzed the efficiency of photoassociation of
an atomic condensate into its molecular counterpart using various
theoretical schemes. Others predicted that the ground state of the
hybrid system would have a soliton-like nature
\cite{DrummondPRL,soliton} with liquid-like properties performing
Josephson-like oscillations in response to a sudden variation of a
magnetic field \cite{Timmermans}. Most frequently, coherent
oscillations between the atomic and molecular condensates are
envisaged \cite{Timmermans,HeinzenPRL,Piza}. All the approaches
employed so far have used a 2-mode approximation (only the
condensates involved) to describe the dynamics of a coupled
ultracold atomic-molecular system.

In this Letter we demonstrate that the 2-mode approach is
inadequate as a description of current experiments on stimulated
production of cold molecules in atomic Bose-Einstein condensates.
For typical parameters the predicted oscillations between the two
condensates are strongly damped due to significant population of
excited atomic and molecular modes leading to a complete depletion
of the initial condensate on a short time scale. The method we
use, quantum-optical in spirit, can be regarded as a generalization of the
classic Bogolubov approximation \cite{Fetter}.

The second-quantized Hamiltionian for the hybrid atomic-molecular
system confined in a box with periodic boundary conditions may be
written in the following form:
\begin{eqnarray}\label{H1}
H&=&\int {\rm d^3} r
\left[\Phi^{\dagger}\frac{p^2}{2m}\Phi+\Psi^{\dagger}\frac{p^2}{4m}\Psi
\right]
\\ &+&\sqrt{V}\hbar\Omega \int {\rm
d^3} r \left[\Psi^{\dagger}\Phi^2 +\Psi\Phi^{\dagger 2}\right]
\nonumber
\\ &+& \frac{V \hbar g}{2} \int {\rm d^3} r \; \Phi^{\dagger}
\Phi^{\dagger} \Phi \Phi \nonumber \; ,
\end{eqnarray}
where $\Phi$ and $\Psi$ are atomic and molecular field operators,
respectively, $V=L^3$ is a volume of the system ($L$ being a size
of the box), $\Omega$ parameterizes a coupling between the two
fields and $g=\frac{4\pi\hbar a}{m V}$ characterizes the atom-atom
interactions in the low-energy, s-wave approximation ($a$ being
the scattering length and $m$ -- the mass of the atom; in fact,
$a$ is slightly changed in external, e.g. optical, fields -- in
present experiments this correction is small though
\cite{HeinzenPRL}). A similar Hamiltonian is used to describe a
process of second-harmonic generation in nonlinear optics
\cite{shg}. The atom-molecule and molecule-molecule collisions are
not included as their parameters in the low-energy regime are
unknown. However, as can be seen from the construction of the
method, their incorporation could be easily accomplished. Note
that in this model a molecule is created when the positions of two
atoms exactly coincide (for an example of a finite-range coupling
approach see \cite{Kostrun}). The fields $\Phi$ and $\Psi$ are
expanded in natural modes of the system -- the plane waves:
\begin{eqnarray}\label{expan}
\Phi({\bf r})=\frac{1}{\sqrt{V}}\sum_{{\bf k}}\exp(-i{\bf k}
\cdot{\bf r})a_{{\bf k}} \ , \\ \Psi({\bf
r})=\frac{1}{\sqrt{V}}\sum_{{\bf k}}\exp(-i{\bf k} \cdot{\bf
r})b_{{\bf k}} \ ,\nonumber
\end{eqnarray}
where $a_{{\bf k}}$ and $b_{{\bf k}}$ are bosonic annihilation
operators for atoms and molecules, respectively, and ${\bf
k}=\frac{2\pi}{L}{\bf n}$ with $n_{i}=0,\pm 1,\pm 2,\ldots$
($i=x,y,z$). With this substitution the Hamiltonian assumes its
final form:
\begin{eqnarray}
\frac{H}{\hbar}&=&\xi \sum_{{\bf k}}n^2(a^{\dagger}_{{\bf
k}}a_{{\bf k}} +\frac{1}{2}b^{\dagger}_{{\bf k}}b_{{\bf k}})
\\ &+&\Omega \sum_{{\bf k},{\bf k'}}
b_{{\bf k}+{\bf k'}}^{\dagger}a_{{\bf k}}a_{{\bf k'}} +b_{{\bf
k}+{\bf k'}}a^{\dagger}_{{\bf k}}a^{\dagger}_{{\bf k'}}+ \nonumber
\\ &+&\frac{1}{2}g
\sum_{{\bf k},{\bf k'},{\bf k''}} a^{\dagger}_{{\bf k}+{\bf
k'}-{\bf k''}} a^{\dagger}_{{\bf k''}}a_{{\bf k'}}a_{{\bf k}}
\nonumber \; ,
\end{eqnarray}
where $\xi=\frac{\hbar}{2m}(\frac{2\pi}{L})^2$. After elimination
of a fast time dependence with the substitution $\{a,b\}_{{\bf
k}}=\exp(-i\xi n^2 t) \{\alpha,\beta\}_{{\bf k}}$, the Heisenberg
equations of motion for the operators $\alpha_{{\bf k}}$ and
$\beta_{{\bf k}}$ acquire the following form:
\begin{eqnarray}\label{Heisenberg}
&& \dot{\alpha_{{\bf k}}} = -2 i \Omega \sum_{{\bf k'}}
\exp(\frac{1}{2}i\xi |{\bf n}-{\bf n'}|^2 t)\beta_{{\bf k}+{\bf
k'}}\alpha^{\dagger}_{{\bf k'}}- \\ && ig \sum_{{\bf k'},{\bf
k''}} \exp\left[2 i\xi ({\bf n}-{\bf n'})\cdot ({\bf n}-{\bf n''})
t\right]\alpha^{\dagger}_{{\bf k'}+{\bf k''}-{\bf k}}\alpha_{{\bf
k'}}\alpha_{{\bf k''}} \nonumber \; , \\ && \dot{\beta_{{\bf k}}}
= - i \Omega \sum_{{\bf k'}} \exp(-\frac{1}{2}i\xi |{\bf n}-2{\bf
n'}|^2 t)\alpha_{{\bf k}-{\bf k'}}\alpha_{{\bf k'}} \nonumber \; .
\end{eqnarray}

\begin{figure}[hbp]
\begin{center}
\leavevmode
\includegraphics[scale=0.6]{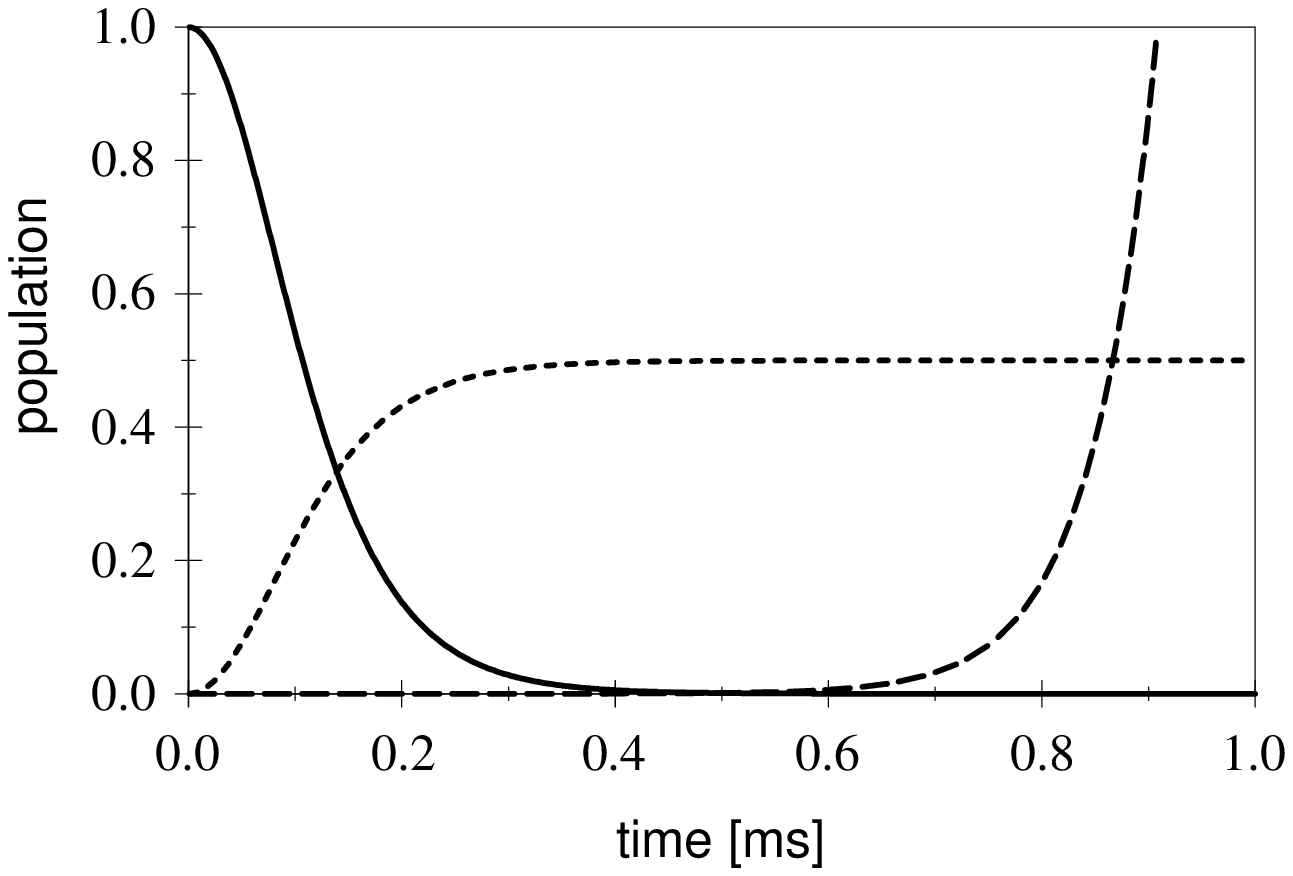}
\caption{Populations of the atomic (solid line) and molecular
(dotted line) condensates calculated within a 2-mode model (see
Eqs.~(\ref{2mode})) and of the quantum corrections (dashed line)
in the 1st excited atomic mode (see Eqs.~(\ref{corrections})). The
parameters are $\Omega=$18.432 Hz, $\xi$=71.373 Hz and $N=10^5$.}
\label{diverge}
\end{center}
\end{figure}

The first hint about limitations of the 2-mode model comes from
the following argument. As initially only the atomic condensate
(the ${\bf k}=0$ mode) is populated, one might naively suspect
that the coupling would primarily lead to an interconversion
between atoms and molecules
\cite{Juha,DrummondPRL,Timmermans,HeinzenPRL,Piza}. In such a case
only the atomic and molecular condensate (${\bf k}=0$) modes would
be macroscopically populated and therefore we replace the
corresponding operators ($\alpha_{0}$ and $\beta_{0}$) by
c-numbers (classical complex fields) and set all the others to
zero ($\alpha_{{\bf k}\neq 0}=0$ and $\beta_{{\bf k}\neq 0}=0$).
Then, in the absence of atomic collisions, time evolution of the
amplitudes can be calculated analytically. Assuming
$\alpha_{0}=r\exp(i\phi)$ and $\beta_{0}=\rho\exp(i\theta)$ with
$\phi=\phi_{0}=const$ and $\theta=\theta_{0}=const$, the solution
is:
\begin{eqnarray}\label{2mode}
\alpha_{0}(t)&=&\frac{\sqrt{N}}{\cosh(\sqrt{2N}\Omega t)}\exp(i
\phi_{0}) \; , \\
\beta_{0}(t)&=&-i\sqrt{\frac{N}{2}}\tanh(\sqrt{2N}\Omega t)
\exp(2i \phi_{0}) \; , \nonumber \\
\theta_{0}&=&2\phi_{0}-\frac{\pi}{2} \; , \nonumber
\end{eqnarray}
where $N$, the total number of atoms, is a conserved quantity
($N=|\alpha_{0}(t)|^2+2|\beta_{0}(t)|^2$) -- see
Fig.~\ref{diverge}. Note that the inclusion of interactions makes
the 2-mode problem analytically insoluble (in particular the
condition of constant phases $\theta$ and $\phi$ cannot be
fulfilled any more). Numerical solutions are of oscillatory
character \cite{Timmermans,HeinzenPRL,Piza} -- different from
(\ref{2mode}). They depict an interconversion between atomic and
molecular condensate modes.

In the next step we calculate quantum corrections to such a 2-mode
model treating the condensates ($\alpha_{0}$ and $\beta_{0}$) as
the only sources of particles. This amounts to leaving in
Eqs.(\ref{Heisenberg}) the terms with at least one ${\bf k}=0$ (in
the form of (\ref{2mode})) and neglecting the other ones. In the
resulting equations only the zero-momentum as well as the ${\bf
k}$ and $-{\bf k}$ atomic and molecular modes are present which
allows to solve the coupled linear operator equations numerically.
Their asymptotics ($t\rightarrow\infty$) may be investigated
analytically though, yielding the following expectations values
for the number operators:
\begin{eqnarray}\label{corrections}
\langle\beta^{\dagger}_{{\bf k}}\beta_{{\bf k}}\rangle&=&const \;
, \\ \langle\alpha^{\dagger}_{{\bf k}}\alpha_{{\bf
k}}\rangle&=&\frac{N\Omega^2}{2\lambda^2}\left[\exp(2\lambda
t)+\exp(-2\lambda t)-2\right] \nonumber \; ,
\end{eqnarray}
where $\lambda=\sqrt{2N\Omega^2-\xi^2 n^4}$. $\lambda$ is real for
all modes with $n<\sqrt[4]{2 N \Omega^2/\xi^2}$, which sets the number
of modes whose population grows in time. As typically both
$\Omega$ and $\xi$ are of the order of $10-10^2$ Hz while
$N\sim10^5-10^6$ \cite{DrummondPRL,HeinzenPRL,HeinzenWWW},
corrections to the 2-mode model are divergent for many low-lying
states. With the parameters used it is only for the 11-th and higher excited
atomic
modes that the quantum corrections are small and oscillatory (imaginary
$\lambda$). Therefore, one is not
allowed to exclude excited modes (${\bf k}\neq0$) from a
theoretical model \cite{Holland}.
\noindent Fig.~\ref{diverge} presents an example of the run-away correction
to the 2-mode model obtained numerically together with
the classical source terms $\alpha_{0}(t)$ and $\beta_{0}(t)$
(note: in all plots the total populations are normalized by the
total number of particles $N$). From Fig.~\ref{diverge} one immediately
concludes that the 2-mode solution (\ref{2mode}) is physically
invalid for times larger than $\sim 0.6$ ms. To cure this problem,
from now on, we will use a multi-mode model in the form of
Eqs.~({\ref{Heisenberg}).

Solution of the nonlinear operator equations (\ref{Heisenberg})
presents an extremely difficult task. A semiclassical
approximation, however, is well justified for all except extremely
low temperatures. Therefore, we replace all operators by c-number
complex amplitudes. From the viewpoint of the Bogolubov method
\cite{Fetter}, such an approach is legitimate as indeed many modes
are macroscopically populated (i.e. their occupation is greater
than quantum fluctuations). This way we are left with a set of
nonlinear differential equations which must be solved numerically.

The first observation in the multi-mode model is that if one
starts from a pure atomic condensate (the ${\bf k}=0$ mode), the
2-mode dynamics is recovered. However, even a very small
occupation of excited atomic or molecular modes results in the
dynamics beyond the 2-mode approach. Such a behavior resembles an
initiation of superfluorescence where quantum, not thermal,
fluctuations play a role. In a typical experiment, roughly $85$\%
of the total number of atoms populate the atomic condensate
whereas the rest of them is thermally distributed over excited
atomic modes and we mimic such a situation in the initial
conditions of our model \cite{comment}. All the molecular modes
are initially unpopulated. Each atomic mode is assigned an
initial, randomly chosen, phase. Any subsequent dynamics depends
on the initial phases and, in a sense, a single simulation
describes a single experimental realization. Values of the
parameters in the model are $\Omega=$18.432 Hz, $\xi=$71.373 Hz,
$N=10^5$ and $g=$0.018 Hz (the atomic mass and the scattering
length are those of $^{87}$Rb and the size of the box is equal to
the Thomas-Fermi radius of a condensate of $N$ atoms in a trap
with frequency of $\omega_{0}=2\pi \; 80$ Hz -- see
\cite{HeinzenPRL,HeinzenWWW}). Sample results for the model with 2622 modes
(maximum atomic and molecular excitation $n_{i}= -5,\ldots,5$,
$i=x,y,z$) are presented in Fig.~\ref{traj}. In fact the results stabilize
for the number of modes exceeding 1000.
\begin{figure}[hbp]
\begin{center}
\leavevmode
\includegraphics[scale=0.6]{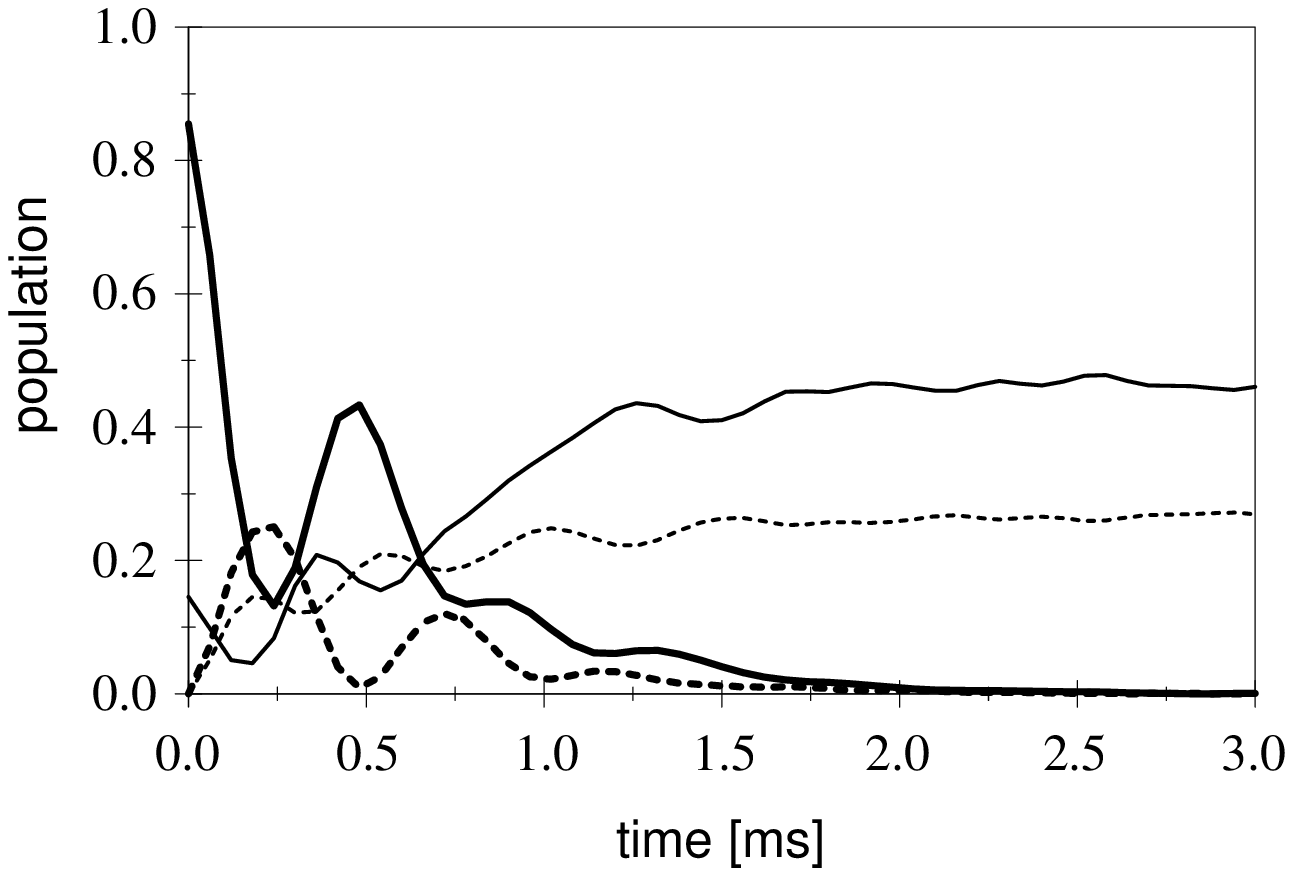}
\caption{Populations of atomic (thick solid line) and molecular
(thick dashed line) condensates (the ${\bf k}=0$ modes) and the
sums of populations of all excited (${\bf k}\neq 0$) atomic (thin
solid line) and molecular (thin dashed line) modes in the model
with 2622 states.}
    \label{traj}
  \end{center}
\end{figure}

A striking discovery is that after a relatively short time both
the atomic and molecular condensates (the ${\bf k}=0$ modes) are
completely depleted and all particles occupy excited modes in
roughly equal proportions. For the used (typical) parameters, only
one oscillation in the condensate population survives -- see
Fig.~\ref{2-multi}.
\begin{figure}[hbp] \begin{center}
\leavevmode
\includegraphics[scale=0.6]{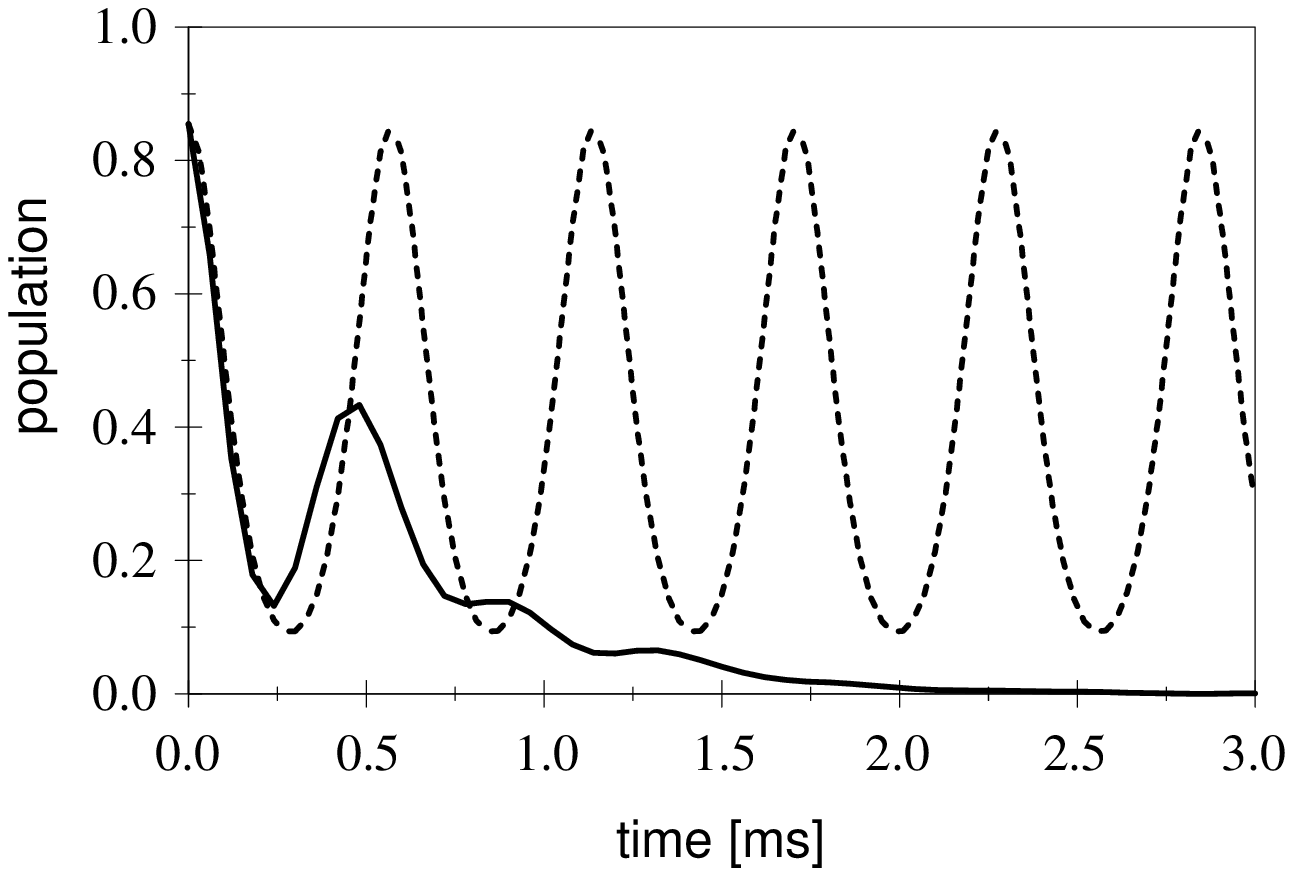}
\caption{2-mode (dashed line) vs. multi-mode (solid line) dynamics
of an atomic condensate population.} \label{2-multi}
\end{center}
\end{figure}
\noindent This clearly indicates that if one wants to convert an atomic
condensate to its molecular counterpart, it is necessary to
precisely tailor the length of the coupling pulse so that the
maximum of the molecular ${\bf k}=0$ mode is picked. If one
investigates the condensate (${\bf k}=0$) population and a sum of
populations of all excited modes (${\bf k}\neq 0$) solely (like in
Fig.~\ref{traj}), their dependence on the initial phases is
negligible in a sufficiently big model (we checked this fact by
setting different initial phases in our calculations). The effect
is due to self-averaging caused by very many random phases present
in the system. However, it can be observed in the time evolution
of single modes. We emphasize that, contrary to the previous
treatments \cite{Timmermans,HeinzenPRL,Piza,HeinzenWWW}, the
effective losses from the condensates are completely due to a
Hamiltonian evolution and not because of any phenomenologically
introduced loss processes. In other words, due to an external weak
coupling of many degrees of freedom, the system is effectively
heated and its final equilibrium state certainly does not result
from any $T=0$ dynamics. The inclusion of atom-molecule and
molecule-molecule collisions would randomize the still coherent
dynamics leading to an analogous and even more pronounced effect.
Remarkably, the character of the dynamics does not depend on the
total number of particles -- our simulations for $N=2\cdot10^3$
and $N=5\cdot10^5$ still show only one oscillation in the atomic
condensate population on a slightly altered (longer for smaller
$N$) time scale. The amplitude of this oscillation is bigger for
smaller $N$ indicating a small increase in the molecular
condensate production ($\sim 30\%$ for $N=2\cdot10^3$ and $\sim
23\%$ for $N=5\cdot10^5$ instead of $\sim 25\%$ for $N=10^5$ as
seen in Fig.~\ref{traj}).

In order to recover the standard oscillatory dynamics of the
2-mode limit \cite{Timmermans,HeinzenPRL,Piza} within a multi-mode
model, one needs to detune the excited modes. The latter can be achieved by
decreasing the size of the box (and so enlarging a spacing between the
excited modes)
while keeping the total density fixed (in order not to alter the
scattering and coupling parameters). The results for the box whose
volume is $15^3$ times smaller are presented in
Fig.~\ref{2modelim} (the appropriately decreased number of
particles is $N=10^5/15^3 \simeq 30$).
\begin{figure}[hbp]
\begin{center}
\leavevmode
\includegraphics[scale=0.6]{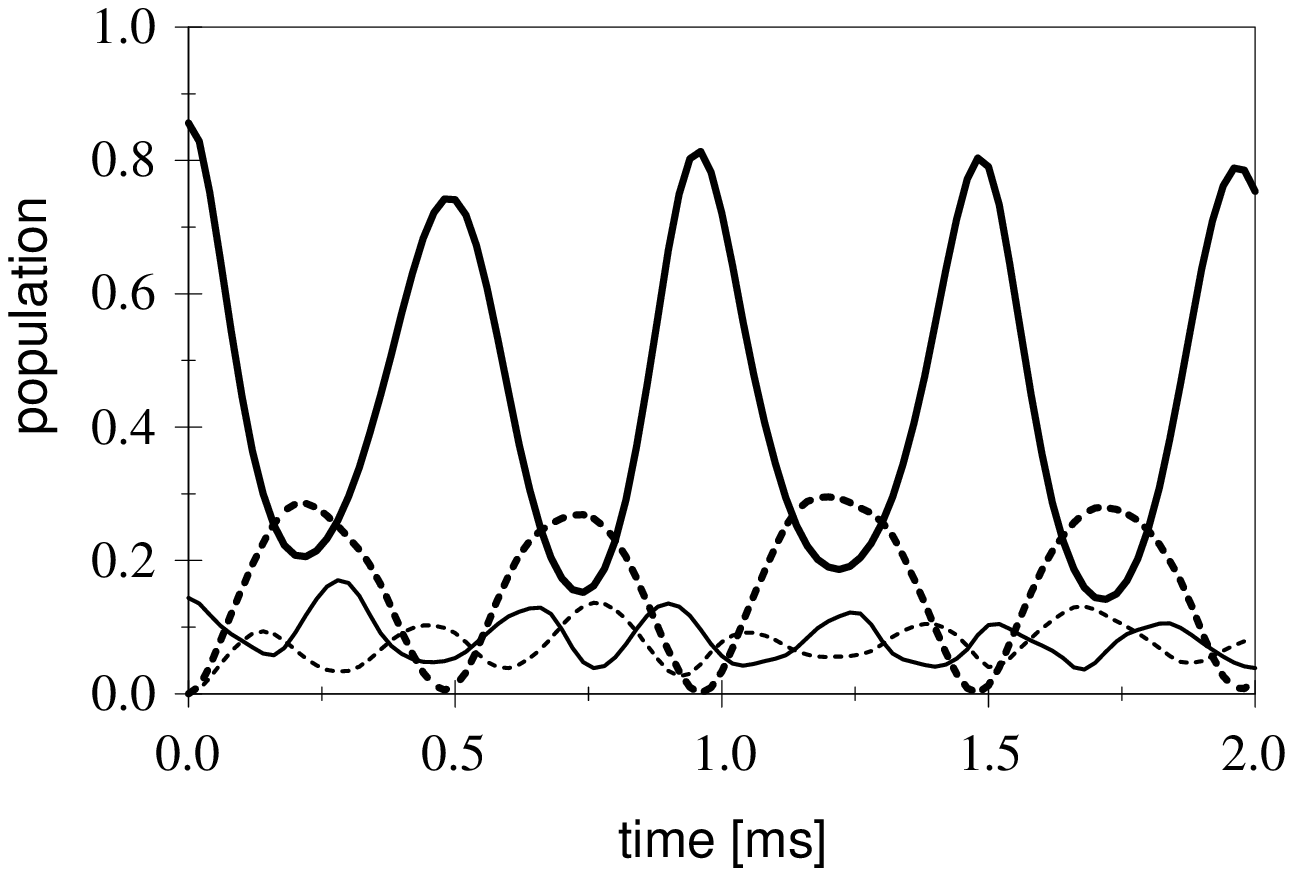}
\caption{Populations of atomic (thick solid line) and molecular
(thick dotted line) condensates (the ${\bf k}=0$ modes) and of the
sums of all excited (${\bf k}\neq0$) atomic (thin solid line) and
molecular (thin dotted line) modes in very small box (see text for
details) -- recovery of the 2-mode dynamics.} \label{2modelim}
\end{center}
\end{figure}
\indent To summarize, we point out the invalidity of a 2-mode
model for a description of present experiments with coupled
atomic-molecular systems in the Bose-Einstein condensation regime.
We find it necessary to employ a multi-mode approach and within it
we observe a complete depletion of both the atomic and molecular
condensates on a short time scale. For typical parameters, only
one giant oscillation between the two condensates is present.
Thus, a new destructive mechanism in the system is pointed out
which should be taken into account while planning future
experiments. An effectively 2-mode dynamics is recovered under
special conditions of large detunings. The system presents an
interesting example of a quantum coherent dynamical evolution
effectively randomized by the coupling of many degrees of freedom.

In this Letter we presented the calculations for the box rather than
the harmonic oscillator potential. We have chosen the box size such that
the first excitation energy is of the order of the level
spacing in the experiment of Heinzen \cite{Heinzen}. Hence, we may expect
that for the harmonic oscillator the time of coherent evolution would be
even shorter, because of the quadratic vs. linear dependence of excitation
energies in the respective potentials.

K.G. and M.G. acknowledge support by Polish KBN grant no 2 P03B 057 15.
K.R. and K.G. are supported by the subsidy of the Foundation for
Polish Science. Part of the results has been obtained using
computers at the Interdisciplinary Centre for Mathematical and
Computational Modeling (ICM) at Warsaw University.


\begin{references}

\bibitem{BEC_exp}
M.H. Anderson {\it et al.}, Science {\bf 269}, 198 (1995); K.B.
Davis {\it et al.}, Phys. Rev. Lett. {\bf 75}, 3969 (1995); C.C.
Bradley {\it et al.}, {\it ibid.} {\bf 75}, 1687 (1995); {\bf 79},
1170(E) (1997); D.G. Fried {\it et al.}, {\it ibid.} {\bf 81},
3811 (1998).

\bibitem{fermions}
B. DeMarco and D.S. Jin, Science {\bf 285}, 1703 (1999).

\bibitem{molecules}
J.D. Weinstein {\it et al.}, Nature (London) {\bf 395}, 148
(1998); A. Fioretti {\it et al.}, Phys. Rev. Lett. {\bf 80}, 4402
(1998); T. Takekoshi, B.M. Patterson, and R.J. Knize, {\it ibid.}
{\bf 81}, 5105 (1998); A.N. Nikolov {\it et al.}, {\it ibid.} {\bf
82}, 703 (1999);  H.L. Bethlem, G. Berden, and G. Meijer, {\it
ibid.} {\bf 83}, 1558 (1999); A.N. Nikolov {\it et al.}, {\it
ibid.} {\bf 84}, 246 (2000); C. Gabbanini {\it et al.}, {\it
ibid.} {\bf 84}, 2814 (2000); H.L. Bethlem {\it et al.}, Nature
(London) {\bf 406}, 491 (2000).

\bibitem{Heinzen} R. Wynar {\it et al.}, Science {\bf 287}, 1016
(2000).

\bibitem{Feshbach}
F.A. van Abeelen and B.J. Verhaar, Phys. Rev. Lett. {\bf 83}, 1550
(1999).

\bibitem{Juha}
J. Javanainen and M. Mackie, Phys. Rev. A {\bf 58}, R789 (1998);
{\bf 59}, R3186 (1999);
M. Mackie and J. Javanainen, {\it ibid.} {\bf 60}, 3174 (1999); M.
Mackie, R. Kowalski, and J. Javanainen, Phys. Rev. Lett. {\bf 84},
3803 (2000).

\bibitem{DrummondPRL}
P.D. Drummond, K.V. Kheruntsyan, and H. He, Phys. Rev. Lett. {\bf
81}, 3055 (1998).

\bibitem{soliton}
K.V. Kheruntsyan and P.D. Drummond, Phys. Rev. A {\bf 58}, R2676
(1999); {\bf 61}, 063816 (2000).

\bibitem{Timmermans}
E. Timmermans {\it et al.}, Phys. Rev. Lett. {\bf 83}, 2691
(1999).

\bibitem{HeinzenPRL}
D.J. Heinzen, R.H. Wynar, P.D. Drummond, and K.V. Kheruntsyan,
Phys. Rev. Lett. {\bf 84}, 5029 (2000).

\bibitem{Piza}
A.N. Salgueiro, M.C. Nemes, M.D. Sampaio, and A.F.R. de Toledo
Piza, quant-ph/9809003.

\bibitem{Fetter}
A.L. Fetter and J.D.Walecka, {\it Quantum Theory
of Many-Particle Systems} (McGraw-Hill, New York, 1991).

%\bibitem{Zoller}
%C.W. Gardiner, P. Zoller, R.J. Ballagh, and M.J. Davis, Phys. Rev.
%Lett. {\bf 79}, 1793 (1997); C.W. Gardiner {\it et al.}, {\it ibid.}
%{\bf 81}, 5266 (1998).

\bibitem{shg}
P.N. Butcher and D. Cotter, {\it The Elements of Nonlinear Optics}
(Cambridge University Press, Cambridge, 1990).

\bibitem{Kostrun}
J. Javanainen and M. Ko\v{s}trun, Opt. Express {\bf 5}, 188
(1999).

\bibitem{HeinzenWWW}
D.J. Heinzen's presentation at the CTAMOP Workshop on
Bose-Einstein Condensation and Degenerate Fermi Gases, February
1999, University of Colorado,
http://condon.colorado.edu/\verb+~+chg/Talks/Heinzen/

\bibitem{Holland}
At the time of writing this manuscript we found out that the
problem of going beyond the 2-mode dynamics had been also
addressed in M. Holland, J. Park, and R. Walser, cond-mat/0005062.
The authors point out an effect of condensate depletion due to a
thermal cloud. However, in their treatment they neglect all
collisions.

\bibitem{comment}
We do recover the 2-mode dynamics in the semiclassical approximation 
starting from the pure atomic condensate (very low temperature). However, as 
we see in Fig.~\ref{diverge}, even in the case of a 100\% initial
occupation of the atomic condensate, the quantum initiation makes the
2-mode model invalid after $\sim 0.6$ ms. 

\end{references}
\end{document}